\documentclass[prd,onecolumn,nofootinbib,showkeys]{revtex4}
\usepackage{bm,amsmath,amssymb,graphicx}
\newcommand\gothfamily{\usefont{U}{ygoth}{m}{n}}
\DeclareTextFontCommand{\textgoth}{\gothfamily}

\begin{document}

\title{PRIMORDIAL FLUCTUATIONS OF SCALE FACTOR IN CLOSED UNIVERSE IN EINSTEIN--CARTAN GRAVITY}

\author{{\bf Nikodem Pop{\l}awski}}

\affiliation{Department of Mathematics and Physics, University of New Haven, 300 Boston Post Road, West Haven, CT 06516, USA}
\email{NPoplawski@newhaven.edu}

\noindent
{\em Modern Physics Letters A}\\
Vol. {\bf 33}, No. 40 (2018) 1850236\\
\copyright\,World Scientific Publishing Company
\vspace{0.4in}

\begin{abstract}
We consider a homogeneous and isotropic Universe, described by the minisuperspace Lagrangian with the scale factor as a generalized coordinate.
We show that the energy of a closed Universe is zero.
We apply the uncertainty principle to this Lagrangian and propose that the quantum uncertainty of the scale factor causes the primordial fluctuations of the matter density.
We use the dynamics of the early Universe in the Einstein--Cartan theory of gravity with spin and torsion, which eliminates the big-bang singularity and replaces it with a nonsingular bounce.
Quantum particle production in highly curved spacetime generates a finite period of cosmic inflation that is consistent with the Planck satellite data.
From the inflated primordial fluctuations, we determine the magnitude of the temperature fluctuations in the cosmic microwave background, as a function of the numbers of the thermal degrees of freedom of elementary particles and the particle production coefficient, which is the only unknown parameter.
\end{abstract}


\maketitle

\section{Introduction}

The observed anisotropies in the temperature of the cosmic microwave background radiation (CMB), discovered by the Cosmic Background Explorer (COBE) and measured more accurately by the Wilkinson Microwave Anisotropy Probe (WMAP), provide precious information about the dynamics of the early Universe \cite{Ric}.
These anisotropies originate from the primordial fluctuations in the density of matter in the early Universe and are the seeds of the cosmic structure that we observe today.
The origin of the primordial fluctuations is attributed to quantum fluctuations of a hypothetical scalar field (inflaton) that also is hypothesized to generate cosmic inflation \cite{infl1,infl2,infl3}.
In \cite{ApJ}, we showed that inflation can be generated by torsion and quantum particle production in highly curved spacetime, so the inflaton is not necessary.

In this paper, we show that the primordial fluctuations can also be generated without the inflaton.
We propose that they can originate from quantum fluctuations of the scale factor in the minisuperspace description of the Universe.
In this description, a homogeneous and isotropic Universe as a whole is represented by a Lagrangian in which the scale factor is the only generalized coordinate \cite{point1,point2,point3,point4,Dal}.
The dynamics of the scale factor is given by the Lagrange equations of motion.
In the minisuperspace formalism, quantum fluctuations of the scale factor can be obtained from the uncertainty relation derived from the canonical commutation relation for the scale factor and the corresponding generalized momentum.
Since the scale factor is a function of the cosmic time, quantum fluctuations of the scale factor in this formalism physically arise from quantum fluctuations of time.
They produce fluctuations in space: different points in space having different values of the scale factor perturb homogeneity and isotropy of the Universe.

The paper is structured as follows.
Section II briefly describes the cosmology of the early Universe with torsion.
Section III reviews the minisuperspace Lagrangian for the Universe with the scale factor as a generalized coordinate.
Section IV shows an interesting result: the energy of a closed Universe calculated from that Lagrangian is zero.
Section V derives the uncertainty principle for the scale factor.
Section VI presents the main results of this paper.
It proposes that the primordial fluctuations of the matter density originate from the quantum uncertainty of the scale factor.
Then it derives the order of magnitude of the resulting temperature fluctuations in the CMB.
The results are briefly summarized in Section VII.

\section{Cosmology in Einstein--Cartan gravity}

The simplest mechanism generating both a nonsingular bounce (instead of the singular big bang) and inflation, involving only one unknown parameter and no hypothetical fields, arises in the Einstein--Cartan (EC) theory of gravity \cite{ApJ}.
EC is the simplest and most natural theory of gravity with torsion, with the Lagrangian density for the gravitational field proportional to the Ricci scalar, as in general relativity.
The conservation law for the total (orbital plus spin) angular momentum of fermions in curved spacetime, consistent with the Dirac equation, requires that the antisymmetric part of the affine connection (torsion tensor) \cite{Schr} is not constrained to zero \cite{req1,req2}.
Instead, torsion is determined by the field equations obtained from varying the action with respect to the torsion tensor \cite{EC1,EC2,EC3,EC4,EC5,EC6,EC7}.
In EC, torsion is coupled to the spin of fermions.
As a result, fermions must be spatially extended \cite{non1,non2,non3}.
The multipole expansion of the conservation law for the spin tensor in EC gives a spin tensor which describes fermionic matter as a spin fluid (ideal fluid with spin) \cite{NSH}.
The effective energy density and pressure of a spin fluid are given by
\begin{equation}
\tilde{\epsilon}=\epsilon-\alpha n_\textrm{f}^2,\quad\tilde{p}=p-\alpha n_\textrm{f}^2,
\label{eff}
\end{equation}
where $\epsilon$ and $p$ are the thermodynamic energy density and pressure, $n_\textrm{f}$ is the number density of fermions, and $\alpha=\kappa(\hbar c)^2/32$ \cite{HHK}.
If we assume that the Universe is homogeneous and isotropic, then it is described by the Friedmann--Lema\^{i}tre--Robertson--Walker (FLRW) metric in the isotropic spherical coordinates \cite{LL2}:
\begin{equation}
ds^2=c^2 dt^2-\frac{a^2(t)}{(1+kr^2/4)^2}(dr^2+r^2 d\theta^2+r^2\sin^2\theta d\phi^2),
\label{metric}
\end{equation}
where $a(t)$ is the scalar factor as a function of the cosmic time $t$, and $k$ is 0 (flat Universe), 1 (closed Universe), or -1 (open Universe).
The Einstein field equations for this metric become the Friedmann equations:
\begin{equation}
\frac{\dot{a}^2}{c^2}+k=\frac{1}{3}\kappa\epsilon a^2
\label{energy}
\end{equation}
and
\begin{equation}
\frac{\dot{a}^2+2a\ddot{a}}{c^2}+k=-\kappa pa^2,
\label{Fri}
\end{equation}
where a dot denotes the derivative with respect to $t$ and $\kappa=8\pi G/c^4$.
The two Friedmann equations give the first law of thermodynamics for an adiabatic Universe, which can be used instead of the second Friedmann equation:
\begin{equation}
\frac{d}{dt}(\epsilon a^3)+p\frac{d}{dt}(a^3)=0.
\label{cont}
\end{equation}
For EC, the Friedmann equations have the same form but the energy density and pressure are replaced by $\tilde{\epsilon}$ and $\tilde{p}$ \cite{KG1,KG2,tor1,tor2,tor3}.

The spin fluid in the early Universe is formed by an ultrarelativistic matter in kinetic equilibrium, for which $\epsilon=h_\star T^4$, $p=\epsilon/3$, and $n_\textrm{f}=h_{n\textrm{f}}T^3$, where $T$ is the temperature of the Universe, $h_\star=(\pi^2/30)(g_\textrm{b}+(7/8)g_\textrm{f})k_\textrm{B}^4/(\hbar c)^3$, and $h_{n\textrm{f}}=(\zeta(3)/\pi^2)(3/4)g_\textrm{f}k_\textrm{B}^3/(\hbar c)^3$ \cite{Ric}.
For standard-model particles, $g_\textrm{b}=29$ and $g_\textrm{f}=90$.
In the presence of spin and torsion, the first Friedmann equation is therefore \cite{ApJ,tor3}
\begin{equation}
\frac{{\dot{a}}^2}{c^2}+k=\frac{1}{3}\kappa(h_\star T^4-\alpha h_{n\textrm{f}}^2 T^6)a^2.
\label{nikoeq1}
\end{equation}
The first law of thermodynamics (\ref{cont}), with $\tilde{\epsilon}$ and $\tilde{p}$, gives \cite{ApJ}
\begin{equation}
\Bigl(\frac{\dot{a}}{a}+\frac{\dot{T}}{T}\Bigr)\Bigl(1-\frac{3\alpha h_{n\textrm{f}}^2}{2h_\star}T^2\Bigr)=0,
\end{equation}
which yields
\begin{equation}
\frac{\dot{a}}{a}+\frac{\dot{T}}{T}=0.
\label{aT}
\end{equation}
Quantum particle production \cite{Zel1,Zel2,Zel3,Zel4,Zel5,Zel6,Prig} caused by extremely high curvatures occurring near a bounce creates enormous amounts of matter and entropy.
Dominating particles are massive spin-1 bosons of the weak nuclear interaction \cite{Zel7}.
The first law of thermodynamics (\ref{aT}) in the presence of particle production becomes \cite{ApJ}
\begin{equation}
\frac{\dot{a}}{a}+\frac{\dot{T}}{T}=\frac{cK}{3h_{n1}T^3},
\label{nikoeq2}
\end{equation}
where $K=a^{-3}d(a^3 n_1)/(cdt)$ is the particle production rate, $h_{n1}=(\zeta(3)/\pi^2)g_{n1}k_\textrm{B}^3/(\hbar c)^3$, and $g_{n1}=10$ is the number of the thermal degrees of freedom of the massive bosons.
Equations (\ref{nikoeq1}) and (\ref{nikoeq2}) describe the dynamics of the early Universe.

Since the negative term on the right-hand side of (\ref{nikoeq1}) scales with $T$ faster ($\sim T^6$) than the positive term ($\sim T^4$), $\dot{a}$ becomes zero at some value of $T=T_\textrm{max}$.
The corresponding value of the scale factor is positive \cite{ApJ}.
The scale factor is never zero and the singularity in the Universe is avoided, as a result of gravitational repulsion at extremely high densities (about $10^{45}$ kg/m$^3$ and higher) generated by the coupling between the spin of fermions and torsion in EC \cite{avert1,avert2,avert3,avert4,tor1,tor2,tor3}.
Consequently, the Universe started expanding from a nonsingular state with a positive minimum scale factor and finite energy density.
Before that state, the Universe was contracting: that state was a bounce.
The origin of the bounce is the negative term on the right-hand side of (\ref{nikoeq1}).
The contracting phase could have originated in a black hole existing in another universe \cite{ApJ,Pat1,Pat2,Pat3,Pat4,Pat5,Pat6,Pat7,Pat8,Pat9}.

Particle production should vanish at a bounce, otherwise the temperature at that instant could exceed $T_\textrm{max}$ which would contradict (\ref{nikoeq1}).
The simplest production rate is given by \cite{ApJ}
\begin{equation}
K=9\beta\Bigl(\frac{H}{c}\Bigr)^4,
\label{rate}
\end{equation}
where $H=\dot{a}/a$ is the Hubble parameter and $\beta$ is a dimensionless particle production coefficient.
A similar form of the particle production rate, proportional to a power of $H$, was proposed in \cite{Mod1,Mod2,Zel8}.
Particle production occurs mostly immediately after a bounce, when $H$ reaches enormously high values.
The Universe expands and its temperature decreases.
Torsion and particle production weaken and the Universe becomes radiation-dominated.
Eventually, the matter in the Universe becomes nonrelativistic and the energy density starts scaling as $a^{-3}$.

Eqautions (\ref{nikoeq2}) and (\ref{rate}) give \cite{ApJ}
\begin{equation}
\frac{\dot{a}}{a}\Bigl[1-\frac{3\beta}{c^3 h_{n1}T^3}\Bigl(\frac{\dot{a}}{a}\Bigr)^3\Bigr]=-\frac{\dot{T}}{T}.
\label{ineq}
\end{equation}
Equations (\ref{nikoeq1}) and (\ref{ineq}) determine the dynamics of the early Universe with torsion and particle production.
The signs of $\dot{a}$ and $\dot{T}$ must be opposite to avoid an indefinite increase of the scale factor (eternal inflation).
Thus, during an expanding phase, the second term in the square bracket in (\ref{ineq}) must be lesser than 1.
Accordingly, the maximum of this term, which can be determined by (\ref{nikoeq1}), must be lesser than 1.
This condition determines an upper limit, $\beta_\textrm{cr}$, for the particle production coefficient \cite{ApJ}.
For $\beta=0$, the closed Universe is oscillatory with an infinite number of bounces and crunches (cycles) \cite{ApJ,SD}.
If $0<\beta<\beta_\textrm{cr}$, then the universe has a finite number of cycles before becoming matter-dominated (nonrelativistic).
As $\beta$ increases, the number of cycles decreases \cite{SD}.

If $\beta$ is slightly lesser than $\beta_\textrm{cr}$, then the Universe has only one bounce \cite{SD}.
The maximum value of the second term in the square bracket in (\ref{ineq}) is slightly lesser than 1.
In this case, (\ref{ineq}) at that maximum value gives $\dot{T}\approx 0$ and $\dot{a}/a=H\approx$ constant.
Accordingly, the Universe has a finite period of a nearly exponential expansion (inflation), $a(t)\sim e^{Ht}$, at a nearly constant energy density \cite{ApJ}.
EC with particle production can thus explain inflation without a scalar field and reheating.
Contrary to the scalar-field models of inflation in which a scalar field causes inflation and then decays into matter, quantum particle production near a bounce creates matter and causes inflation which ends when torsion becomes weak:  when $\alpha h_{n\textrm{f}}^2 T^6$ in (\ref{nikoeq1}) becomes negligible compared to $h_\star T^4$.
Depending on the particle production rate, such a Universe may undergo one or more several nonsingular bounces \cite{ApJ,SD}.
The last bounce, after which the Universe expands to the size where dark energy starts dominating, can be regarded as the big bounce that replaces the big bang.

\section{Lagrangian for Universe}

Since EC and particle production can naturally avoid the initial singularity and explain inflation, the inflaton field is not necessary.
Accordingly, the primordial fluctuations in standard cosmology which are attributed to quantum fluctuations of the inflaton, should be explained by another mechanism.
We propose that the primordial fluctuations originated from quantum fluctuations of the scale factor, which arise from quantum fluctuations of time since $a$ is a function of $t$.
To investigate such a scenario, we must describe the Universe as a whole, in terms of the Lagrangian with the scale factor as a generalized coordinate.

We consider a closed, homogeneous, and isotropic Universe, $k=1$.
We keep $k$ in equations for clarity.
The action for the gravitational field and matter is given by
\begin{equation}
S=\frac{1}{c}\int\Bigl(-\frac{R}{2\kappa}\Bigr)\sqrt{-g}d\Omega+S_\textrm{m},
\label{action}
\end{equation}
where $S_\textrm{m}=\int L_\textrm{m}dt$ is the action for matter, $L_\textrm{m}$ is the Lagrangian for matter, $g$ is the determinant of the metric tensor $g_{ij}$, and $d\Omega$ is an element of four-volume \cite{LL2}.
For the metric (\ref{metric}), the Ricci scalar is equal to
\begin{equation}
R=-\frac{6}{a^2}\Bigl(\frac{\dot{a}^2+a\ddot{a}}{c^2}+k\Bigr).
\label{Ricci}
\end{equation}
Substituting (\ref{Ricci}) into (\ref{action}) and integrating over space (which reduces to multiplying the integrand by the volume of a closed Universe \cite{LL2} $V=2\pi^2 a^3$) gives $S=\int L\,dt$, where $L$ is the Lagrangian for the gravitational field and matter:
\begin{equation}
L=\frac{6\pi^2}{\kappa}\Bigl(\frac{a\dot{a}^2+a^2\ddot{a}}{c^2}+ka\Bigr)+L_\textrm{m}.
\label{total}
\end{equation}
Substituting $a^2\ddot{a}=(d/dt)(a^2\dot{a})-2a\dot{a}^2$ into (\ref{total}) and omitting a total time derivative, which does not contribute to the equations of motion \cite{LL1}, gives \cite{Dal}
\begin{equation}
L=\frac{6\pi^2}{\kappa}\Bigl(\frac{-a\dot{a}^2}{c^2}+ka\Bigr)+L_\textrm{m}.
\label{Lag}
\end{equation}
This Lagrangian is time-reparametrization invariant \cite{HK}.
The action for this Lagrangian is thus
\begin{equation}
S=\frac{6\pi^2}{\kappa}\int\Bigl(\frac{-a\dot{a}^2}{c^2}+ka\Bigr)dt+S_\textrm{m}.
\label{act}
\end{equation}
The Lagrangian (\ref{Lag}), known as the Lagrangian in the minisuperspace approximation \cite{point1,point2,point3,point4,Dal}, describes a homogeneous and isotropic Universe as a whole.
It determines the dynamics of the scale factor in a symmetric spacetime, coupled to a matter source.

Varying the action (\ref{act}) with respect to $a$ and $g_{ij}$ gives
\begin{equation}
\delta S=\frac{6\pi^2}{\kappa}\int\Bigl(\frac{-\dot{a}^2}{c^2}+k\Bigr)\delta a\,dt-\frac{12\pi^2}{\kappa}\int\frac{a\dot{a}}{c^2}\delta\dot{a}\,dt-\frac{1}{2c}\int T^{ij}\delta g_{ij}\sqrt{-g}d\Omega,
\end{equation}
where $T^{ij}$ is the energy--momentum tensor for matter \cite{LL2}.
Using $a\dot{a}\,\delta\dot{a}=a\dot{a}\,d\delta a/dt=(d/dt)(a\dot{a}\,\delta a)-(\dot{a}^2+a\ddot{a})\delta a$ and omitting a total time derivative leads to
\begin{eqnarray}
& & \delta S=\frac{6\pi^2}{\kappa}\int\Bigl(\frac{-\dot{a}^2}{c^2}+k\Bigr)\delta a\,dt+\frac{12\pi^2}{\kappa}\int\frac{\dot{a}^2+a\ddot{a}}{c^2}\delta a\,dt-\frac{1}{2c}\int T^{ij}\frac{\partial g_{ij}}{\partial a}\delta a\sqrt{-g}d\Omega \nonumber \\
& & =\frac{6\pi^2}{\kappa}\int\Bigl(\frac{-\dot{a}^2}{c^2}+k\Bigr)\delta a\,dt+\frac{12\pi^2}{\kappa}\int\frac{\dot{a}^2+a\ddot{a}}{c^2}\delta a\,dt-\frac{1}{2}\int 2\pi^2 a^3 T^{ij}\frac{\partial g_{ij}}{\partial a}\delta a\,dt.
\end{eqnarray}
This variation vanishes for arbitrary variations $\delta a$, so the sum of the integrands must be equal to zero, leading to
\begin{equation}
\frac{\dot{a}^2+2a\ddot{a}}{c^2}+k=\frac{\kappa}{6}a^3 T^{ij}\frac{\partial g_{ij}}{\partial a}.
\label{tens}
\end{equation}
For the metric (\ref{metric}), only the spatial coordinates contribute to $\partial g_{ij}/\partial a$ as follows:
\begin{equation}
\frac{\partial g_{\alpha\beta}}{\partial a}=\frac{2}{a}g_{\alpha\beta}.
\end{equation}
The matter in the Universe can be approximated as an ideal fluid as follows:
\begin{equation}
T^{ij}=(\epsilon+p)u^i u^j-pg^{ij},
\label{fluid}
\end{equation}
where $\epsilon=\epsilon(a)$ is the energy density of matter, $p=p(a)$ is its pressure, and $u^i$ is its four-velocity.
In the comoving frame of reference, in which the four-velocity of the fluid satisfies $u^0=1$ and $u^\alpha=0$, (\ref{tens}) becomes
\begin{equation}
\frac{\dot{a}^2+2a\ddot{a}}{c^2}+k=-\frac{\kappa}{3}a^2 pg^{\alpha\beta}g_{\alpha\beta}=-\kappa pa^2,
\label{sec}
\end{equation}
which is the second Friedmann equation (\ref{Fri}).

Although the calculations in this section used $k=1$, the Lagrangian (\ref{Lag}) is also valid for a flat or an open Universe.

\section{Energy of closed Universe}

One can show that the Lagrangian for matter is equal to \cite{HK}
\begin{equation}
L_\textrm{m}=-\epsilon V=-2\pi^2 \epsilon a^3. 
\end{equation}
The Lagrangian (\ref{Lag}) becomes
\begin{equation}
L=\frac{6\pi^2}{\kappa}\Bigl(\frac{-a\dot{a}^2}{c^2}+ka-\frac{1}{3}\kappa\epsilon a^3\Bigr).
\label{Lagra}
\end{equation}
The Lagrange equations \cite{LL1}, equivalent to the principle of least action, for the scale factor $a$ as a generalized coordinate are \cite{Dal}
\begin{equation}
\frac{d}{dt}\frac{\partial L}{\partial\dot{a}}=\frac{\partial L}{\partial a}.
\end{equation}
They give
\begin{equation}
\frac{1}{c^2}\frac{d}{dt}(-2a\dot{a})=-\frac{\dot{a}^2}{c^2}+k-\kappa\epsilon a^2-\frac{1}{3}\kappa a^3 \frac{\partial\epsilon}{\partial a}.
\label{equat}
\end{equation}
Using
\begin{equation}
\frac{\partial\epsilon}{\partial a}=\frac{\partial(U/V)}{\partial V}\frac{\partial V}{\partial a}=\Bigl(V^{-1}\frac{\partial U}{\partial V}-\frac{U}{V^2}\Bigr)\frac{\partial V}{\partial a}=-\frac{p+\epsilon}{V}\frac{\partial V}{\partial a}=-3\frac{p+\epsilon}{a},
\label{derivative}
\end{equation}
where $U=\epsilon V$ is the internal energy of matter, and substituting it into (\ref{equat}) gives the second Friedmann equation (\ref{Fri}) \cite{Dal}, proving that $L_\textrm{m}=-\epsilon V$.
The Lagrangian for matter acts as a potential energy (equal to the internal energy) contributing to the Lagrangian.

The energy of the Universe is given by
\begin{equation}
E=\frac{\partial L}{\partial\dot{a}}\dot{a}-L=\frac{12\pi^2}{\kappa}\Bigl(\frac{-a\dot{a}^2}{c^2}\Bigr)-\frac{6\pi^2}{\kappa}\Bigl(\frac{-a\dot{a}^2}{c^2}+ka-\frac{1}{3}\kappa\epsilon a^3\Bigr)=\frac{6\pi^2}{\kappa}\Bigl(\frac{-a\dot{a}^2}{c^2}-ka+\frac{1}{3}\kappa\epsilon a^3\Bigr)
\label{totalen}
\end{equation}
and is constant because the Lagrangian does not depend explicitly on time \cite{LL1}.
This constancy means that the Universe as a whole can be regarded as an isolated mechanical system.
Using the first Friedmann equation (\ref{energy}) shows that the total energy (\ref{totalen}) of the gravitational field and matter in a closed Universe is zero.
This is an original derivation, in the minisuperspace description, of the result in \cite{universe1,universe2,universe3} that was obtained by means of the energy--momentum pseudotensors.
Equivalently, one can derive the first Friedmann equation from the condition that the energy of a closed Universe be zero.

\section{Uncertainty principle for scale factor}

Having described the dynamics of the Universe in terms of the Lagrangian with the scale factor as a generalized coordinate, we can now proceed to the uncertainty prinicple applied to this dynamics.
In a quantum theory, Hamilton's principle of least action \cite{LL2,LL1} is generalized to Schwinger's variational principle \cite{Schw1,Schw2,Schw3}, according to which the variation of the transition amplitude between an initial state $|\alpha_i\rangle$ and a final state $|\alpha_f\rangle$ is equal to $i/\hbar$ times the matrix element connecting the two states of the variation $\delta S$ of the action integral $S$:
\begin{equation}
\delta\langle\alpha_f|\alpha_i\rangle=\frac{i}{\hbar}\langle\alpha_f|\delta S|\alpha_i\rangle.
\end{equation}
For any operator $O$ in the Heisenberg picture, this principle gives
\begin{equation}
\delta O=-\frac{i}{\hbar}[O,\delta S],
\label{princ}
\end{equation}
where square brackets denote a commutator and $\delta S$ is the variation of the action at the boundary of the integration domain used to calculate the action.
From Schwinger's principle, one can derive the canonical commutation relation for a generalized coordinate operator and the corresponding (conjugate) generalized momentum operator:
\begin{equation}
[q_i,p_j]=i\hbar \delta_{ij},
\label{relation}
\end{equation}
which gives the Heisenberg uncertainty principle for these two observables:
\begin{equation}
\Delta q_i \, \Delta p_j \geq \frac{\hbar}{2} \delta_{ij},
\end{equation}
where $\Delta$ denotes the standard deviation.

We apply the commutation relation for the Universe, treating the scale factor $a$ as a generalized coordinate operator $q$.
Using the Lagrangian (\ref{Lag}), the generalized momentum corresponding to the scale factor is \cite{HK}
\begin{equation}
p_a=\frac{\partial L}{\partial\dot{a}}=-\frac{12\pi^2 a\dot{a}}{\kappa c^2}.
\end{equation}
Equation (\ref{relation}), $[a,p_a]=i\hbar$, therefore gives
\begin{equation}
\Bigl[a,-\frac{12\pi^2 a\dot{a}}{\kappa c^2}\Bigr]=-\frac{12\pi^2}{\kappa c^2}a[a,\dot{a}]=i\hbar
\end{equation}
or
\begin{equation}
[a,\dot{a}]=-\frac{i\hbar\kappa c^2}{12\pi^2 a}.
\end{equation}
This equation is the commutation relation for the scale factor $a$ and its time derivative $\dot{a}$.
The uncertainty principle corresponding to this commutation relation is
\begin{equation}
\Delta a\,\Delta\dot{a}\geq\frac{\hbar\kappa c^2}{24\pi^2 a}.
\label{uncer}
\end{equation}

A closed Universe is mathematically equivalent to the three-dimensional hypersurface of a four-dimensional hypersphere whose radius is the scale factor.
Fluctuations of the scale factor represent deviations of the hypersurface from a hyperspherical shape.
Although the scale factor is a macroscopic quantity, it can be regarded as a local variable at each point in the Universe that determines the local values of the energy density and pressure of matter, as noted below (\ref{fluid}).
Therefore, the relation (\ref{uncer}) has a physical meaning of an uncertainty principle for the local energy density (or pressure).

\section{Primordial fluctuations}

Each point in the Universe has a local value of the uncertainty of the scale factor $\Delta a$.
We propose that the uncertainty $\Delta a$ in a homogeneous and isotropic (on average) Universe causes the primordial fluctuations of the matter density (related to the fluctuations of the temperature), which make the Universe inhomogeneous on smaller scales.
In (\ref{nikoeq1}), the variable $a$ can be written as a mean value $\bar{a}$ plus a perturbation $\delta a$: $a=\bar{a}+\delta a$.
Similarly, $T=\bar{T}+\delta T$.
Substituting these expressions for $a$ and $T$ into (\ref{nikoeq1}), omitting terms of higher order than linear in $\delta a$ and $\delta T$, and subtracting (\ref{nikoeq1}) written for $a=\bar{a}$ and $T=\bar{T}$, gives (after dropping the bars for convenience):
\begin{equation}
3\dot{a}\,\delta\dot{a}-c^2\kappa a(h_\star T^4-\alpha h_{n\textrm{f}}^2 T^6)\delta a-c^2\kappa a^2(2h_\star T^3-3\alpha h_{n\textrm{f}}^2 T^5)\delta T=0.
\label{pert}
\end{equation}
Henceforth, $a$ and $T$ will denote the mean values of the scale factor and temperature for the whole Universe.
In this equation, we can substitute
\begin{equation}
\delta\dot{a}\sim\frac{\hbar\kappa c^2}{24\pi^2 a\,\delta a},
\end{equation}
following the uncertainty relation (\ref{uncer}).

The Hubble length $l_\textrm{H}=c/H$ gives the range of causal interaction in cosmological timescales.
As the Universe enters inflation, the Hubble length decreases.
Perturbations greater than $l_\textrm{H}$ freeze in comoving coordinates \cite{Ric} (this is called the horizon exit) until the Hubble length increases again (in the radiation-dominated or matter-dominated era) and exceeds their size.
Consequently, the local uncertainties $\delta a$ freeze when $l_\textrm{H}$ decreases below $\delta a$.
The Universe becomes composed of many causally disconnected regions evolving with different local values of $a$.
We will focus here on determining the value of $\delta a$ at the instant of the horizon exit.
Accordingly, we put in (\ref{pert})
\begin{equation}
\dot{a}\sim\frac{ca}{\delta a}.
\end{equation}
Furthermore, we consider the case where the particle production coefficient $\beta$ is slightly lesser than $\beta_\textrm{cr}$.

During inflation, the second term in the square bracket in (\ref{ineq}) is slightly lesser than 1.
Accordingly, we can estimate that
\begin{equation}
\frac{3\beta}{h_{n1}T^3}\Bigl(\frac{\dot{a}}{ca}\Bigr)^3\approx 1.
\label{limit}
\end{equation}
Equation (\ref{pert}) gives
\begin{equation}
(h_\star T^4-\alpha h_{n\textrm{f}}^2 T^6)\frac{3\beta a}{h_{n1}T^3}+(2h_\star T^3-3\alpha h_{n\textrm{f}}^2 T^5)\Bigl(\frac{3\beta}{h_{n1}}\Bigr)^{2/3}\frac{a^2}{T^2}\delta T\approx\frac{\hbar c}{8\pi^2},
\label{deltaT}
\end{equation}
which determines the order of magnitude of $\delta T/T$ as a function of $T$ and $a$ at the horizon exit and $\beta$.
The scale factor at the horizon exit is given by
\begin{equation}
\Bigl(\frac{h_{n1}}{3\beta}\Bigr)^{2/3}T^2+\frac{k}{a^2}\approx\frac{1}{3}\kappa(h_\star T^4-\alpha h_{n\textrm{f}}^2 T^6),
\label{scalef}
\end{equation}
which results from (\ref{nikoeq1}) and (\ref{limit}).
This value of the scale factor estimates the characteristic length of curvature below which quantum-mechanical effects are significant.
After the horizon exit, the right-hand side of (\ref{uncer}) decreases exponentially, quantum effects can be neglected, and comovingly frozen perturbations begin to evolve classically.
The duration of the period from the big bounce to the horizon exit is given by the integral $\int da/\dot{a}$ from the scale factor at the big bounce \cite{ApJ} to its value given by (\ref{scalef}).

Finally, the temperature at the horizon exit is on the order of the temperature at which inflation occurs \cite{ApJ}, which is
\begin{equation}
T=\Bigl(\frac{h_\star}{2\alpha h_{n\textrm{f}}^2}\Bigr)^{1/2}.
\label{temp}
\end{equation}
This temperature can be obtained by finding the maximum of the right-hand side of (\ref{nikoeq1}).
Equations (\ref{deltaT}), (\ref{scalef}), and (\ref{temp}) determine the order of magnitude of $\delta T/T$ at the horizon exit as a function of $\beta$, provided $\beta$ is slightly lesser than $\beta_\textrm{cr}$ (which is the condition for inflation in the proposed scenario).
Conversely, the observed order of magnitude of $\delta T/T \sim 2\times 10^{-5}$ can determine the value of $\beta$.

\section{Conclusions}

We showed that the primordial fluctuations, caused by the uncertainty principle applied to the positively curved Universe as a whole, may be the origin of the observed order of magnitude of the temperature fluctuations in the CMB.
We used the EC theory of gravity which extends general relativity by taking into account the spin of fermions, resulting in torsion.
Spin and torsion in EC generate gravitational repulsion at extremely high densities.
This repulsion eliminates the big-bang singularity and replaces it with a bounce.
Quantum particle production in highly curved spacetime immediately after a bounce can generate a finite period of inflation which ends when torsion becomes weak.
Contrary to scalar-field models of inflation in which a scalar field causes inflation and then decays into matter, quantum particle production creates matter and causes inflation.
Accordingly, hypothetical scalar fields are not necessary to explain the dynamics of the early Universe.

To explain the primordial fluctuations that lead to temperature fluctuations in CMB, we proposed that they originate from the uncertainty principle for the scale factor.
To implement this principle, we described the Universe in terms of the Lagrangian with the scale factor as a generalized coordinate.
The equations of motion for this system become the Friedmann equations.
We showed that the energy of the closed Universe is zero.
We applied Schwinger's variational principle to the Lagrangian of the Universe and proposed that the quantum uncertainty of the scale factor $\Delta a$, arising from the quantum uncertainty of the cosmic time, causes the primordial fluctuations of the matter density.
The subsequent dynamics of these fluctuations is determined by the dynamics of the scale factor, however, we focused on $\delta T/T$ at the horizon exit.
The exponential expansion of the Universe with torsion and quantum particle production (for $\beta$ slightly lesser than $\beta_\textrm{cr}$) immediately after the big bounce \cite{ApJ} predicts the CMB parameters: the scalar spectral index of density perturbations, its running, and the tensor-to-scalar ratio, that are consistent with the Planck 2015 observations \cite{Planck2015}, as was shown in \cite{SD}.
The dynamics of the early Universe driven by torsion and particle production is equivalent to that with a plateau-like scalar-field potential \cite{SD}, which is supported by the Planck 2013 observations \cite{Planck2013}.
Standard power-law scalar-field potentials, favored by the inflaton, which are not supported by those data \cite{ASL}.
Consequently, the spin-torsion explanation of inflation and primordial fluctuations appears to be advantageous.

\section*{Acknowledgment}

This work was funded by the University Research Scholar program at the University of New Haven.

\end{document}